\documentstyle[aps,preprint,epsfig]{revtex}

\begin{document}

\draft

\title{Absence of weak localization in two-dimensional disordered
Frenkel lattices}

\author{A.\ Rodr\'{\i}guez$^{1}$, M.\ A.\ Mart\'{\i}n-Delgado$^{2}$,
J. Rodriguez-Laguna$^{2}$, G. Sierra$^{3}$, V.\ A.\
Malyshev$^{4}$, F.\ Dom\'{\i}nguez-Adame$^{5}$, and J.\ P.\
Lemaistre$^{6}$}

\address{$^{1}$GISC, Departamento de Matem\'{a}tica Aplicada y
Estad\'{\i}stica, Universidad Polit\'{e}cnica, E-28040 Madrid, Spain\\
$^{2}$Departamento de F\'{\i}sica Te\'orica I, Universidad
Complutense, E-28040 Madrid, Spain\\
$^{3}$Instituto de Matem\'aticas y F\'{\i}sica Fundamental,
CSIC, Madrid, Spain\\
$^{4}$National Research Center "Vavilov State Optical
Institute", 199034 Saint-Petersburg, Russia \\
$^{5}$GISC, Departamento de F\'{\i}sica de Materiales,
Universidad Complutense, E-28040 Madrid, Spain\\
$^{6}$Laboratoire des Milieux D\'esordonn\'es et
H\'et\'erog\`enes, Universit\'e P.\ et M.\ Curie, CNRS UMR7603, Paris, France}

\date{\today}

\maketitle

\begin{abstract}

The standard one-parameter scaling theory predicts that all eigenstates in
two-dimensional random lattices are weakly localized. We show that this claim
fails in two-dimensional dipolar Frenkel exciton systems. The linear energy
dispersion at the top of the exciton band, originating from the long-range
inter-site coupling of dipolar nature, yields the same size-scaling law for the
level spacing and the effective disorder seen by the exciton. This finally
results in the delocalization of those eigenstates in the thermodynamic limit.
Large scale numerical simulations allow us to perform a detailed multifractal
analysis and to elucidate the nature of the excitonic eigenstates.

\end{abstract}

\bigskip

\pacs{{\it Keywords}: \quad Frenkel excitons, Anderson  localization,
disordered solids}

One of the most attractive problems in condensed matter physics is the
localization of quasi-particles (electrons, phonons, excitons) in disordered
matter. The existence or absence of the localization-delocalization transition
has been found to strongly depend on the system dimensionality. The
one-parameter scaling theory of localization~\cite{Abrahams79} states that any
nonzero disorder causes exponential localization of all eigenstates in
one-dimensional (1D) and two-dimensional (2D) systems, regardless their
energies, while in three-dimensional (3D) systems only a rather strong disorder
causes the state localization. It is to be noted that there exist several
exceptions to this rule. In this regard, anomalously weak localization is known
to occur at the band center in 1D~\cite{Inui94} and 2D~\cite{Inui94,Eilmes98}
systems with off-diagonal disorder. Moreover, correlations in disorder may
cause delocalization of states in 1D systems~\cite{Flores89,Dunlap90,Bellani99}
even in the presence of strong disorder. The long-range inter-site coupling
also may act as a driving force for delocalization~\cite{Logan87,Levitov89} in
any dimension. We have recently shown the absence of localization in a 1D
Hamiltonian with a special type of long-range intersite interaction, resulting
in a specific, non-parabolic quasi-particle energy dispersion. Remarkably, the
delocalized states belong to one tail of the band~\cite{Rodriguez00}.

In this work we report further progress along the lines mentioned above and
demonstrate that extended states may occur in 2D disordered Frenkel systems,
where the exciton eigenenergies  of the ordered system scale {\em linearly\/}
with the wavenumber at the top of the exciton band. In such a case, the level
spacing decreases on increasing the system size in the same manner that the
strength of {\em effective\/} disorder seen by the exciton [see
Eq.~(\ref{sigma2}) below]. Therefore, if the disorder is of {\em
perturbative\/} magnitude for a given lattice size, it will remain {\em
perturbative\/} on increasing the size, consequently allowing delocalized
eigenstates.

We consider a Frenkel exciton Hamiltonian on a regular ${\cal N} = N\times N$
lattice with diagonal disorder:
\begin{equation}
{\cal H}= \sum_{\bf n}\> \epsilon_{\bf n} |{\bf n}\rangle
\langle{\bf n}| +
\sum_{\bf mn}\>J_{\bf nm}^{} |{\bf n}\rangle \langle{\bf m}|.
\label{Hamiltonian}
\end{equation}
Here, $|{\bf n}\rangle$ is the state vector of the ${\bf n}$-th excited
molecule with energy $\epsilon_{\bf n}$ and ${\bf n} = (n_x,n_y)$, $n_x,n_y$
being integers  ($-N/2 \le n_x,n_y < N/2$, with $N$ even). The intersite
dipole-dipole interaction is taken in the form $J_{{\bf nm}}=J/|{\bf n}-{\bf
m}|^{3}$, where $J>0$ is the coupling between nearest-neighbor (NN) molecules
in the lattice (hereafter we assume that transition dipole moments of molecules
are perpendicular to the plane of the system and that their magnitudes are the
same). The joint distribution function of a realization of disorder is the
product of box functions of width $\Delta$ centered around zero. The quantity
$\Delta/J$ is referred to as {\em degree of disorder}.

In the excitonic representation, assuming periodic boundary conditions, the
Frenkel Hamiltonian~(\ref{Hamiltonian}) takes the form
\begin{mathletters}
\label{1}
\begin{equation}
{\cal H} = \sum_{\bf K}E_{\bf K}^{} |{\bf K}\rangle \langle{\bf
K}| + \sum_{\bf KK^\prime}(\delta{\cal H})_{\bf KK^\prime}^{}
|{\bf K}\rangle \langle{\bf K^\prime}|,
\label{1H}
\end{equation}
where ${\bf K} = (2\pi/N)(k_x,k_y)$ runs over the first Brillouin zone,
$k_x,k_y$ being integers ranging within the interval $-N/2 \le k_x,k_y < N/2$.
Here $E_{\bf K}$ is the unperturbed exciton eigenenergy
\begin{equation}
E_{\bf K}=J\sum_{{\bf n}\neq {\bf 0}} \> {1 \over|{\bf n|^3}}\,e^{i{\bf
K}\cdot{\bf n}},
\label{1Ek}
\end{equation}
and $(\delta {\cal H})_{\bf KK^\prime}$ is the inter-mode coupling matrix
\begin{equation}
(\delta {\cal H})_{\bf KK^\prime}= {1\over{\cal N}}\sum_{{\bf n}}
\epsilon_{\bf n} e^{i({\bf K - K^\prime)\cdot n}}.
\label{1dH}
\end{equation}
\end{mathletters}

Hereafter we keep long-range terms in~(\ref{1Ek}) due to their major role. It
can be shown that near the extreme points of the band ${\bf K} = {\bf 0}$ and
${\bf K} = {\bbox \pi} \equiv (\pi,\pi)$ the exciton energy spectrum takes a
{\it linear} and a parabolic form, respectively~\cite{Christiansen98}:
\begin{mathletters}
\begin{eqnarray}
E_{\bf K} & \simeq & 9.03J - 2\pi J|{\bf K}|, \quad
\quad |{\bf K}|\ll 1,
\label{E_top} \\
E_{\bf K} & \simeq & - 2.65J + 0.4 J|{\bf K} - {\bbox\pi}|^{2},
\quad \quad |{\bf K} - {\bbox\pi}| \ll 1.
\label{E_bottom}
\end{eqnarray}
\end{mathletters}
{}From this it follows that the energy spacing close to the top of the exciton
band behaves as $N^{-1} = {\cal N}^{-1/2}$, while in the vicinity of the bottom
scales as $N^{-2} = {\cal N}^{-1}$.

Depending on the degree of disorder and the lattice size, the operator $\delta
{\cal H}$ may couple the extended excitonic states $|{\bf K}\rangle$ to each
other, thus resulting in their localization. Our task now is to calculate the
typical fluctuation of this matrix in order to gain  insight into the magnitude
of the exciton inter-mode coupling. The corresponding magnitude of interest is
$\sigma_{\bf KK^\prime}^2 = \big\langle |(\delta {\cal H})_{\bf
KK^\prime}|^2\big\rangle$, where the angular brackets indicate the average over
the distribution $\prod_{\bf n} P(\epsilon_{\bf n})$. After performing the
average one gets
\begin{equation}
\sigma_{\bf KK^\prime} \sim \sigma \equiv {\Delta \over \sqrt{\cal N}}.
\label{sigma2}
\end{equation}
Here $\sigma$ is referred to as effective degree of disorder. As we can see,
the typical magnitude of the inter-mode coupling scales as ${\cal N}^{-1/2} =
N^{-1}$. The most remarkable fact is that $\sigma$ decreases on increasing
${\cal N}$ in the same manner as the level spacing close to the top of the
exciton band. This finding has a  dramatic effect on the localization
properties of the states within this region. Indeed, consider for example a
finite lattice of size $N\times N$ and the two first unperturbed states with
${\bf K} = {\bf 0}$ and ${\bf K^\prime} = (2\pi/N,0)$. The energy difference
between them, according to Eq.~(\ref{E_top}), is $\delta E = 4\pi^2 J/N$. Take
now the degree of the disorder to be $\Delta \ll 4\pi^2 J$. Notice that this
condition is not very restrictive since in actual systems the degree of
disorder is expected not to exceed the exciton bandwith ($\simeq 11.68J$).
Under this condition, the strength of the effective disorder $\sigma =
\Delta/N$, governing  the exciton state mixing and thereof localization, is of
{\em perturbative\/} magnitude, namely $\sigma \ll \delta E$. What is most
important, it will remain {\em perturbative\/} upon increasing the lattice size
because both magnitudes scale similarly with $N$. Hence, these states will not
be mixed by disorder and will remain extended over the entire lattice
independent of its size. It is clear that the same conclusion can be drawn for
all the states of the linear spectrum range except the degenerate ones. They
are mixed by any small amount of disorder. However, since different sets of
degenerate states are not coupled to each other due to the {\em perturbative\/}
nature of the effective degree of disorder ($\sigma \ll \delta E$), the above
conclusion is also valid with regard to degenerate states.

Let us now turn to the parabolic range of the energy spectrum, where the level
spacing decreases as ${\cal N}^{-1} = N^{-2}$ upon increasing the lattice size,
i.e., faster than the effective degree of disorder $\sigma$ (the same behavior
takes place for both edges of the band obtained within the NN approximation,
namely taking $J_{\bf nm}=0$ when $|{\bf n}-{\bf m}|>1$). Now, even if one
starts with a {\em perturbative\/} magnitude of $\Delta$ at a fixed lattice
size (so that $\sigma \ll \delta E$), it becomes {\em non-perturbative\/} for
larger sizes, resulting finally in localization of those eigenstates.

The results of numerical diagonalization of the
Hamiltonian~(\ref{Hamiltonian}), by means of the Lanczos method~\cite{Golub96},
unambiguously confirm our qualitative  arguments. To examine the character of
the exciton eigenfunction (localized or extended) we have calculated the
inverse participation ratio (IPR) of the uppermost exciton state, according to
the standard definition $ {\mathrm IPR} = \sum_{\mathbf{n}}\> |\Psi_{\nu \bf
n}|^{4}, $ where the sum runs over lattice sites and it is assumed that the
eigenfunction $\Psi_{\nu \bf n}$ of the $\nu\,$th eigenstate is normalized to
unity.  On increasing the lateral size $N$, the IPR scales as $N^{-2}$ for
delocalized states, spreading uniformly over a 2D system. On the contrary,
localized states exhibit constant values for different $N$.

The IPR as a function of the lateral size $N$ is shown in Fig.~\ref{fig1} when
the dipole-dipole interaction between  all molecules is taken into account as
well as within the NN approximation. The plots comprise the result of $20$
averages over disorder realizations and $\Delta=J$ in all cases. The slope of
the straight line, being equal to $-1.91$ when dipole-dipole interaction is
taken into account, is close to the theoretical value $-2$. This scaling
suggests the fairly extended nature of the uppermost exciton state. Notice that
there is no scaling of the IPR within the NN approximation, in perfect
agreement with the well-known results stating that those exciton states are
localized. Figure~\ref{fig2} shows an increase of the IPR at a threshold value
$\simeq 13J$, suggesting the occurrence of a smooth delocalization-localization
transition.

A comprehensive way to characterize the spatial distribution of eigenfunctions 
is the computation of the singularity spectrum, as explained in, e.g., 
Ref.~\onlinecite{Schreiber96}. If we cover the system with $(M/L)^2$ boxes of
size $L^2$ (in units of the lattice spacing) and define the normalized $q\,$th
moments $\mu_k(q,\delta)=\mu_k^q(\delta)/\sum_{k^\prime}
\mu_{k^\prime}^q(\delta)$ (where $\delta=L/M$) of the probability distribution
$\mu_k(\delta)=\sum_{n\in \mbox{\scriptsize box}\,k}|\Psi_n|^2$ of finding an
exciton in the $k\,$th box, we may calculate the Lipschitz-H\"{o}lder
exponents 
\begin{equation}
\alpha(q)=\lim_{\delta\to0}\sum_k\mu_k(q,\delta)\ln\mu_k(1,\delta)/\ln\delta
\end{equation} 
which take into account the scaling of the content of each  box with the box
size, as well as the corresponding  value of the singularity spectrum 
\begin{equation}
f(q)=\lim_{\delta\to0}\sum_k\mu_k(q,\delta)\ln\mu_k(q,\delta)/\ln\delta.
\end{equation}

The invariance of the singularity spectrum with the system size for a given
value of disorder is usually taken~\onlinecite{Schreiber96} as a proof of the
occurrence of the Anderson transition. Nevertheless strong fluctuations of the
eigenstates of the system near the transition makes it difficult to calculate
the $f(\alpha)$ curve for the threshold value of $\Delta$. We have calculated
the singularity spectrum for values of the degree of disorder below the
threshold. Figure~\ref{fig3} shows the broadening of the $f(\alpha)$ curve with
increasing disorder, indicating that the excitonic eigenfunctions at the top of
the band becomes progressively more and more localized. This result suggests
again the occurrence of a smooth delocalization-localization
transition, as in Fig.~\ref{fig2}.

In summary, we have shown that the statement of the one-parameter scaling
theory~\cite{Abrahams79} about the weak localization in two dimensions, i.e.,
that any amount of disorder results in localization of all eigenstates, fails
near the top of the exciton band where the quasi-particle spectrum scales {\em
linearly\/} with the wavenumber $|\bf K|$. The states lying at such energy
range are delocalized at moderate strength of disorder  and undergo the
continuous Anderson transition as the disorder degree increases. In our
opinion, the failure of the one-parameter scaling theory for the conditions
considered in the present work is due to the fact that this theory deals only
with the size scaling of the energy spacing  but pays no attention to the
subsequent renormalization of the disorder~(\ref{sigma2}). As it follows from
our treatment, the latter effect plays a major role in localization phenomena,
violating the one-parameter scaling and thus leading to the impossibility to
match our results by this theory.

The authors thank E.\ Maci\'{a}, A.\ S\'{a}nchez, E.\ Diez, R.\ R\"{o}mer and
M.\ Hilke for discussions. F.~D-A. and A.~R. were supported by DGI-MCyT
(Project~MAT2000-0734). V.~A.~M. acknowledges support from INTAS (Project
No.~97-10434).

\begin{figure}
\caption{Lateral size scaling of the inverse participation ratio of the
uppermost exciton eigenfunction at $\Delta=J$, obtained by averaging over $20$
realizations of the disorder.}
\label{fig1}
\end{figure}

\begin{figure}
\caption{IPR of the uppermost exciton eigenfunction as a function of the degree
of disorder for a system of size $N\times N$ (shown in the plot). Results
comprise $10$ realizations of disorder.} 
\label{fig2}
\end{figure}

\begin{figure}
\caption{Singularity spectrum $f(\alpha)$ of the uppermost exciton
eigenfunction for various degrees of disorder for a system of size $96\times
96$. Results comprise $5$ realizations of disorder.} 
\label{fig3}
\end{figure}

\newpage

\thispagestyle{empty}

\centerline{Figure~\ref{fig1}}
\begin{figure}
\centerline{\epsfig{file=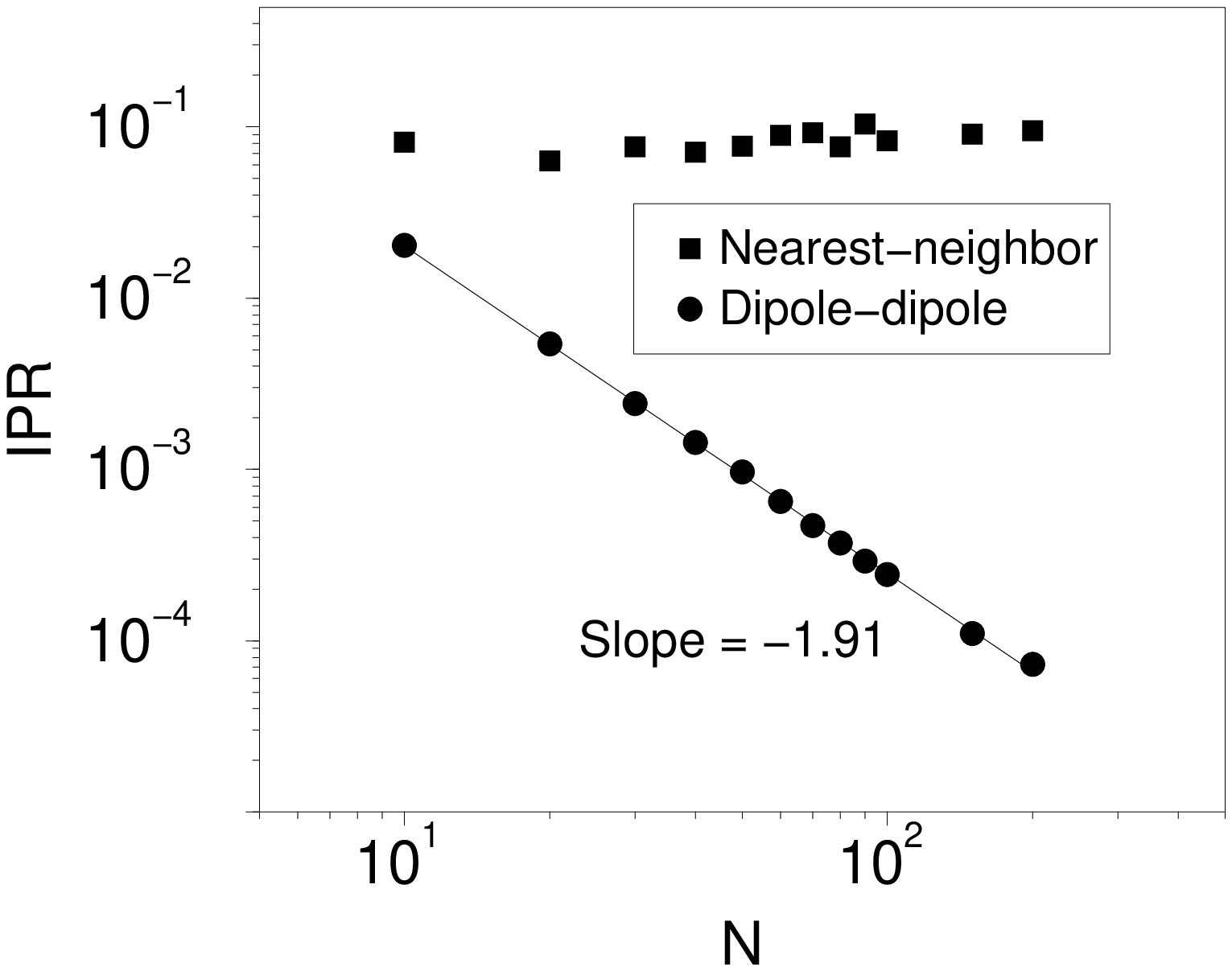,width=7cm}}
\end{figure}

\centerline{Figure~\ref{fig2}}
\begin{figure}
\centerline{\epsfig{file=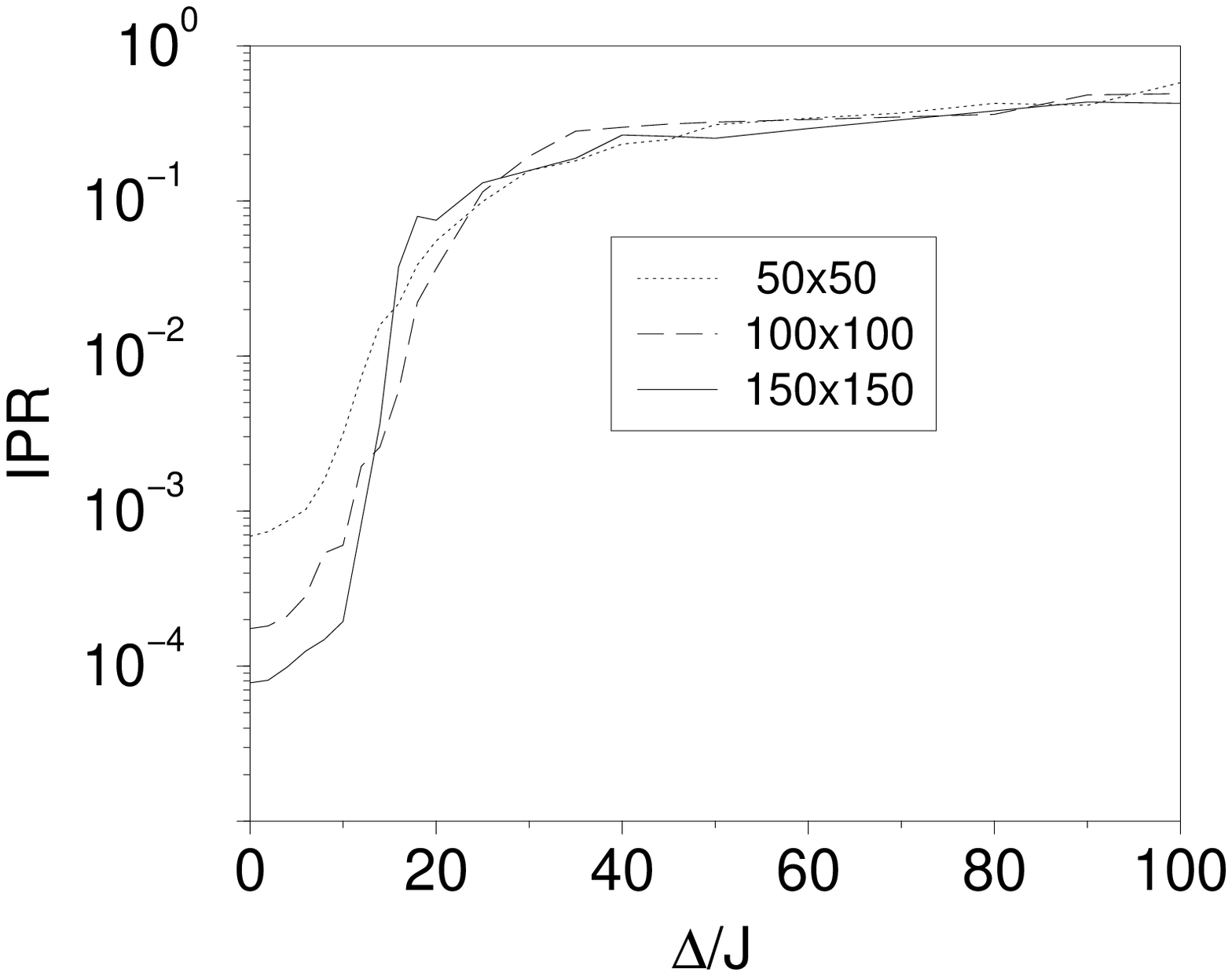,width=7cm}}
\end{figure}

\centerline{Figure~\ref{fig3}}
\begin{figure}
\centerline{\epsfig{file=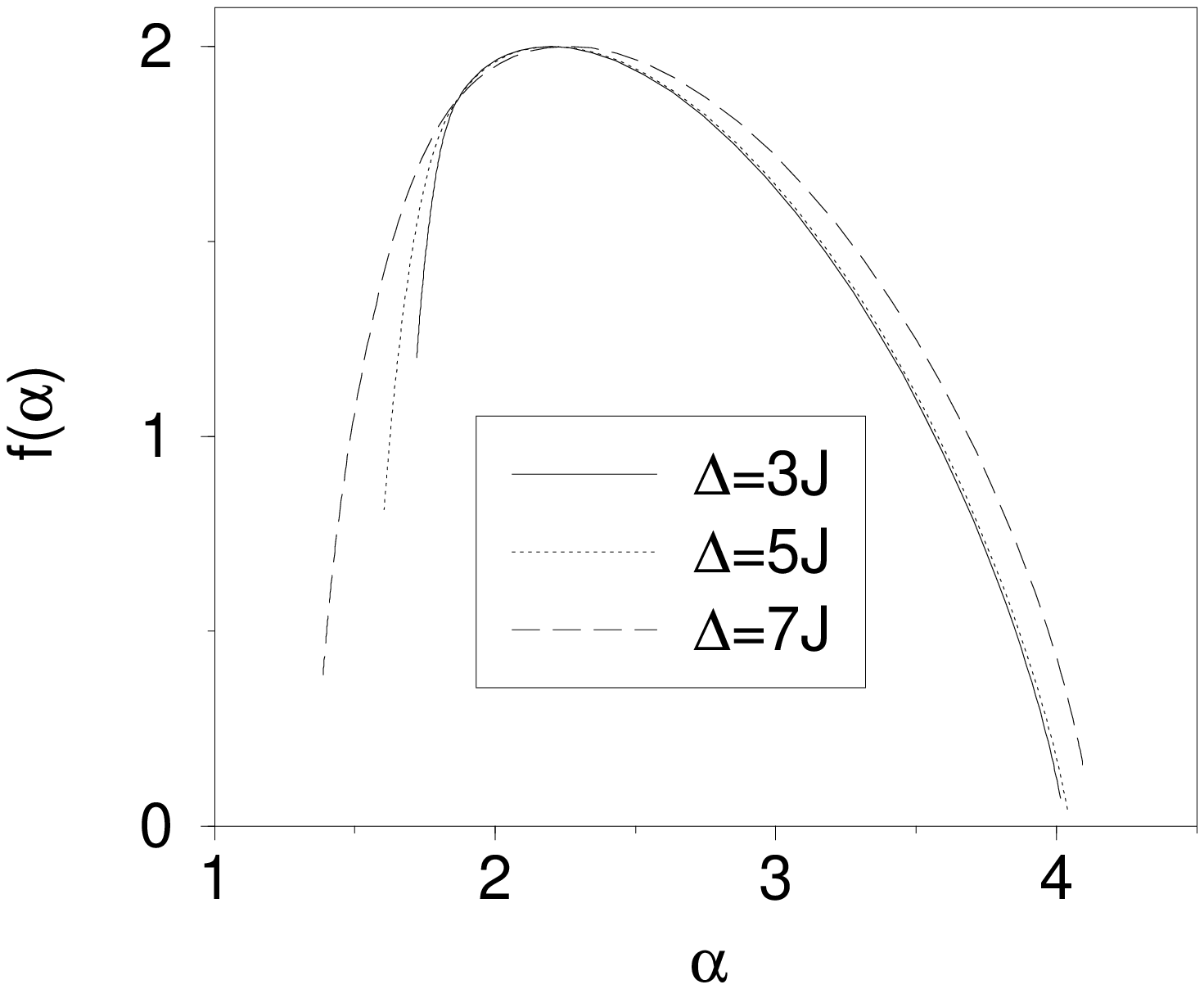,width=7cm}}
\end{figure}

\end{document}